\documentclass[twocolumn,10pt]{article} % here we use the article class, rather than elsarticle

\usepackage[square,numbers,sort&compress,comma]{natbib}

\usepackage{amsmath}
\usepackage{amssymb}
\usepackage{caption}
\usepackage{graphicx}
\usepackage{latexsym}
\usepackage{times}
\usepackage[pagewise]{lineno}

\topmargin - 12pt % might need to be set to 0pt for some installations
\renewenvironment{abstract}%
              {% - begin definition
               \small% - select font
               {\bfseries \abstractname}% - select font
               \par% - end a paragraph (skip \parsep)
               \vspace{10pt}% - add vertical space
              }% - complete definition

\renewcommand\abstractname{Abstract}

\newcommand{\nomenclature}% - name of command
              [1]% - number of arguments
              {% - begin definition
               \bgroup% - begin a local group
               \flushleft% - turn on flushleft option
               \small\bf% - select font
               #1% - insert title text
               \par% - end a paragraph (skip \parsep)
               \egroup% - terminate local group
              }% - complete definition

\renewcommand{\section}% - name of command
              [1]% - number of arguments
              {% - begin definition
               \bgroup% - begin a local group
               \flushleft% - turn on flushleft option
               \small\bf% - select font
               \refstepcounter{section}% - increment counter
               \arabic{section}. #1% - insert title text
               \par% - end a paragraph (skip \parsep)
               \egroup% - terminate local group
              }% - complete definition

\renewcommand{\subsection}% - name of command
              [1]% - number of arguments
              {% - begin definition
               \bgroup% - begin a local group
               \flushleft% - turn on flushleft option
               \small\em% - select font
               \refstepcounter{subsection}% - increment counter
               \arabic{section}.% - insert title text
               \arabic{subsection}. #1% - insert title text
               \par% - end a paragraph (skip \parsep)
               \egroup% - terminate local group
              }% - complete definition

\renewcommand{\subsubsection}% - name of command
              [1]% - number of arguments
              {% - begin definition
               \bgroup% - begin a local group
               \flushleft% - turn on flushleft option
               \small\em% - select font
               \refstepcounter{subsubsection}% - increment counter
               \arabic{section}.% - insert title text
               \arabic{subsection}.% - insert title text
               \arabic{subsubsection}. #1% - insert title text
               \par% - end a paragraph (skip \parsep)
               \egroup% - terminate local group
              }% - complete definition

  \newcommand{\acknowledgement}% - name of command
              [1]% - number of arguments
              {% - begin definition
               \bgroup% - begin a local group
               \flushleft% - turn on flushleft option
               \small\bf% - select font
               #1% - insert title text
               \par% - end a paragraph (skip \parsep)
               \egroup% - terminate local group
              }% - complete definition

  \newcommand{\sectionbib}% - name of command
              [1]% - number of arguments
              {% - begin definition
               \bgroup% - begin a local group
               \flushleft% - turn on flushleft option
               \small\bf% - select font
               #1% - insert title text
               \par% - end a paragraph (skip \parsep)
               \egroup% - terminate local group
              }% - complete definition

\newcommand\DBB{D_{\mathrm{BB}}}
\usepackage{color}
\usepackage{soul}

\begin{document}

\title{\LARGE Large Eddy Simulation of the evolution of the soot size\\
              distribution in turbulent nonpremixed bluff body flames}

\author{{\large Hernando Maldonado Colmán$^{a,*}$, Michael E. Mueller$^{a}$}\\[10pt]
        {\footnotesize \em $^a$Department of Mechanical and Aerospace Engineering, Princeton University, Princeton, NJ 08544, USA}\\[-5pt]
       }

\date{}

% -------------------------------------------------------------------- %
% -------------------------------------------------------------------- %
% -------------------------------------------------------------------- %

\small
\baselineskip 10pt

% -------------------------------------------------------------------- %
% -------------------------------------------------------------------- %
% -------------------------------------------------------------------- %

\twocolumn[\begin{@twocolumnfalse}
\vspace{50pt}
\maketitle
\vspace{40pt}
\rule{\textwidth}{0.5pt}
\begin{abstract} % 100 to 300 words.
Large Eddy Simulation (LES) was used to investigate the evolution of the soot size distribution in a series of turbulent nonpremixed bluff body flames, with different bluff body diameters. The new Bivariate Multi-Moment Sectional Method (BMMSM) is employed to characterize the size distribution. BMMSM combines elements of sectional methods and methods of moments and is capable of reproducing fractal aggregate morphology, thanks to its joint volume-surface formulation, all at relatively low computational costs with fewer transported soot scalars compared to traditional sectional methods. LES results show soot volume fraction profiles agreeing correctly with the experimental measurements, exhibiting significant improvement compared to previous work using the Hybrid Method of Moments (HMOM). The evolution of the particle size distribution function (PSDF) was examined across the flame series and show that the size distribution is less sensitive to the bluff body diameter than the overall soot volume fraction, which increases with increasing bluff body diameter. The PSDF across the flame exhibit different features compared to turbulent nonpremixed jet flames. The long residence times in the recirculation zone leads to a nearly bimodal size distribution, which eventually becomes bimodal in the downstream jet-like region. Further analysis indicates that a size distribution model is needed to correctly predict the soot evolution. Remarkably, due to improved descriptions of oxidation with BMMSM compared to HMOM, significant nucleation and condensation rates in both the recirculation zone and jet-like region were found using BMMSM, on the same order of magnitude as surface growth and oxidation, leading to the improved prediction of mean soot volume fraction compared to HMOM. This work reveals that the need of size distribution is crucial to both predict global soot quantities accurately and reproduce fundamental mechanisms.
\end{abstract}
\vspace{10pt}
\parbox{1.0\textwidth}{\footnotesize {\em Keywords:} Soot; Soot size distribution; Bivariate Multi-Moment Sectional Method; Large Eddy Simulation; Bluff body flames}
\rule{\textwidth}{0.5pt}
\vspace{10pt}

\end{@twocolumnfalse}] 

\clearpage

%\linenumbers

\section{Introduction\label{sec:introduction}} \addvspace{10pt}

The imperative to eliminate soot particles in energy and propulsion systems, limited by regulations now targeting even specific particle sizes, underscores the importance of investigating soot size distribution behavior in turbulent reacting flows. The evolution of the soot size distribution in practical configurations is influenced by complex flow features such as recirculation that control residence time-composition-temperature histories. Fundamental investigations into the evolution of the soot size distributions in such flows requires not only experimental configurations featuring such flows but also computational models for the evolution of the soot size distribution that are accurate yet affordable.

Laboratory-scale configurations have been designed to study recirculating flows, such as swirl flames and bluff body flames, and have been approached computationally. The former includes investigations of the evolution of the soot size distribution using Large Eddy Simulation (LES) \cite{grader2018pressurized,gallen2023investigation}. The latter has received less attention with computational studies limited to mean soot quantities. Joint experimental-computational studies of soot formation in a turbulent nonpremixed bluff body flame configuration were first conducted by Mueller et al. \cite{mueller2013experimental}. This work included experimental observations in the recirculation zone and jet-like region as well as a computational segment focusing on the soot dynamics using LES but constrained to the recirculation zone due to soot subfilter model deficiencies. Deng et al. \cite{deng2017hydrodynamic} conducted similar investigations to understand the impact of adding hydrogen in the fuel stream. Finally, Maldonado Colmán et al. \cite{colman2023largeBBF} studied soot evolution in a new bluff body flame series \cite{rowhani2022effects,rowhani2021soot} using LES. This study introduced a new model for turbulence-chemistry-soot interactions in LES \cite{colman2023largeFRO} that enabled significant soot growth and consumption in the jet-like region, unlike previous studies, allowing for characterization of the mechanisms of the soot evolution in the entire flame, and was also used to validate the LES model versus experimental measurements with variations in the bluff body diameter. However, these studies of bluff body flames used the Hybrid Method of Moments \cite{mueller2009hybrid} for soot, which provides no insight about the soot size distribution.

Recently, a new model that can track the evolution of the soot size distribution in LES of turbulent reacting flows have been developed by Maldonado Colmán et al. \cite{colman2023largeBBF}, called the Bivariate Multi-Moment Sectional Method (BMMSM). It combines elements of both sectional methods and methods of moments. BMMSM includes a joint surface-volume formalism that allows for the consideration of soot's fractal aggregate morphology. Including three moments per section, fewer overall soot scalars are needed compared to traditional sectional models, allowing for both fast and accurate simulations. BMMSM was validated against experiments in both laminar premixed flames and turbulent nonpremixed jet flames.

%Having BMMSM readily available encourages to re-examine the bluff body flame configuration for a computational investigation of the evolution of the soot size distribution in the new flame series, a task that has never been attempted before.
This work aims to leverage the BMMSM soot model in LES of the bluff body flame series \cite{rowhani2022effects,rowhani2021soot}, to analyze the impact of the bluff body diameter, and to elucidate the underlying mechanisms of the evolution of soot size distribution. This manuscript first introduces briefly the LES modeling framework. Then, the bluff body flame configuration and computational setup are presented. Computational results are first validated for soot volume fraction against experimental measurements and compared against results using HMOM from Ref.~\cite{colman2023largeBBF}, followed by a study of the evolution of the soot size distribution in the flame series. Finally, an comprehensive analysis of the mechanism of soot evolution is carried out.

\section{Modeling framework\label{sec:modeling_framework}} \addvspace{10pt}

The LES modeling framework includes combustion, soot, and turbulence-chemistry-soot interaction models, which are presented below. %in the following subsections.

\subsection{Combustion model\label{subsec:combustion}} \addvspace{10pt}

The combustion model is based on the Radiation Flamelet/Progress Variable (RFPV) model expanded for sooting flames~\cite{mueller2012model,yang2019large}. Acquired from steady state nonpremixed flamelet solutions, the thermochemical states are parameterized in terms of mixture fraction $Z$, progress variable $C$, and heat loss parameter $H$. To compensate the mixture's local leaning, due to Polycyclic Aromatic Hydrocarbons (PAH) withdrawal from the gas phase for soot inception~\cite{mueller2012model}, a source term $\dot{m}_{Z}$ is included in $Z$ equation. For the same reason, the source term of $C$ is adapted to take account of the local variation in effective fuel. Finally, both the removal of PAH and radiation losses are considered in the heat loss parameter source term, setting $H=0$ for adiabatic combustion. The radiation modeled is based on an optically thin gray approach for both gas~\cite{barlow2001scalar} and soot~\cite{hubbard1978infrared}. To consider different effective Lewis number of species, a Strain-Sensitive Transport Approach \cite{yang2020large} depending on their characteristic length scales is included. Finally, a lumped PAH transport equation is included \cite{mueller2012model}, which accommodates the slow PAH chemistry with respect to other combustion products included in thermochemical database; the PAH source source term is divided into three parts: chemical production, consumption, and dimerization.

\subsection{Soot model\label{subsec:soot}} \addvspace{10pt}

A statistical representation of the soot particle size distribution considers a Number Density Function (NDF). An NDF with a bivariate formalism is considered to capture fractal aggregate features, by employing soot volume and surface area as internal coordinates \cite{mueller2009joint}. The Population Balance Equation (PBE) governs the NDF spatiotemporal evolution. The recently developed Bivariate Multi-Moment Sectional Method (BMMSM) \cite{colman2023BMMSM} is considered to solve the PBE. The NDF is divided into multiple sections along the volume coordinate to capture the size distribution. Within a section, the NDF is reconstructed by considering a linear distribution as a function of the volume, similar to work of Yang et al. \cite{yang2019multi}. An additional closure to presume the soot volume-surface area dependence is considered, by assuming the section's primary particle diameter is constant, that is, a constant surface area to volume ratio within each section. Then, three statistical moments per section are considered, representing the number density, soot volume fraction, and soot surface area per section. BMMSM accounts for bivariate closure of source terms, including nucleation (from PAH dimerization), condensation (PAH-soot collisions), coagulation (particle-particle collisions), soot surface growth (via acetylene), and oxidation (oxidation-induced fragmentation not included). These were developed based on the previous univariate approach \cite{yang2019multi} and the Hybrid Method of Moments (HMOM) \cite{mueller2009hybrid,mueller2009joint}. This model was previously validated in laminar premixed and turbulent nonpremixed jet flames \cite{colman2023BMMSM}.

\subsection{Turbulence-chemistry-soot interactions\label{subsec:interactions}} \addvspace{10pt}

The governing equations of turbulent sooting flames requires further modeling closures to capture the interactions of turbulence, chemistry, and soot at the LES subfilter scales.

Subfilter turbulence-chemistry interactions are closed using a presumed beta subfilter Probability Density Function (PDF) approach for the mixture fraction. Subfilter turbulence-chemistry-soot interactions are modeled based on the presumed subfilter PDF model from Maldonado Colmán et al.~\cite{colman2023largeFRO} and enables finite-rate oxidation of soot, which was adapted to BMMSM in Ref.~\cite{colman2023BMMSM}. The model considers sooting and non-sooting modes as the presumed bimodal PDF from Ref.~\cite{mueller2011large}. The sooting mode considers a transition from lean to rich mixtures (soot 'off' to soot 'on') that considers the relative rates of soot oxidation and transport to set the sharpness, as opposed to an earlier model with a sharp cutoff that implicitly assumed infinitely fast oxidation \cite{yang2019large}. Finally, and similar to HMOM~\cite{mueller2011large,mueller2012model}, an additional transport equation for the square of the total number density is solved to compute the subfilter intermittency, which mathematically represents the weights between sooting and non-sooting modes. This model was extensively validated in turbulent sooting flames using HMOM \cite{colman2023largeFRO,duvvuri2023relative,colman2023largeBBF} and recently using BMMSM~\cite{colman2023BMMSM}.

%represents soot oxidation as it come across with rich mixtures. The predecessor model from Ref.~\cite{yang2019large} considered a precipitate transition from a soot-enabling mixture to a zero-soot mixture as soon as soot oxidation rates exceeded surface growth rates, so assuming an infinitely fast oxidation strictly controlled by mixing. The new model from Ref.~\cite{colman2023largeFRO} considers instead a gentle transition to represent the effects of finite-rate oxidation of soot compared to local transport rates, which is function of the mixture fraction dissipation rate. 

%Here, manifold solutions from subsection \ref{subsec:combustion} are convolved with the beta PDF, resulting in a new thermochemical database that is stored in a lookup table. The look up table is parameterized in terms of the filtered mixture fraction $\widetilde{Z}$, subfilter mixture fraction variance $Z_{v}$, filtered progress variable $\widetilde{C}$, and filtered heat loss parameter $\widetilde{H}$.

\section{Flame series and computational details\label{sec:details}} \addvspace{10pt}

The flame series configuration and computational setup are briefly summarized in this section.

\subsection{Bluff body flame configuration\label{subsec:bbf}} \addvspace{10pt}

A bluff body flame series consisting of three burners is considered, with similar geometrical features except for their bluff body diameters $\DBB=38$, $50\rm\:mm$, and $64\rm\:mm$. These cases are labeled as ENB-1, ENB-2, and ENB-3, respectively, similar to work of Rowhani and coworkers \cite{rowhani2022effects,rowhani2021soot,rowhani2022relationships}. In all three cases, the central fuel jet has a $4.6\rm\:mm$-diameter, and flow parameters are identical. A 4:1 mixture (by volume) of ethylene and nitrogen is injected at the central jet, with bulk velocity of $32.1\rm\:m/s$ and Reynolds number $\rm Re=15,000$. The coflowing air is injected with bulk velocity of $20\rm\:m/s$. An extensive experimental database is available to compare with simulation results, including observations of flow field using PIV\:\cite{rowhani2022effects} and soot volume fraction using LII~\cite{rowhani2021soot,rowhani2022relationships}. %Refer to work of Rowhani and coworkers~\cite{rowhani2022effects,rowhani2021soot,rowhani2022relationships} for more details on the burners' geometry, flow conditions, and experimental measurements.

\subsection{Computational setup\label{subsec:computational}} \addvspace{10pt}

NGA, a structured finite difference solver for low Mach number turbulent reacting flows, is utilized to carry out the grid-filtered LES computations \cite{desjardins2008high,macart2016semi}. Cylindrical coordinates are considered with identical domains for each flame. Similar to Ref.~\cite{colman2023largeBBF}, the domain's dimensions are $\rm0.96\rm\:m\times0.24\rm\:m$ in the streamwise and radial directions, respectively. The domain is discretized in the streamwise, radial, and circumferential directions with $384\times192\times64$ grid points, respectively. The grid is stretched in both the streamwise and radial directions. A separate simulation of a non-reacting periodic pipe flow with the same bulk velocity is carried out to generate the central jet's inflow condition for each case. The coflow consists of a uniform velocity profile without any turbulent fluctuations. Adiabatic walls are considered at the burner.

Lagrangian dynamic Smagorinsky(-like) models \cite{germano1991dynamic, meneveau1996lagrangian} are employed to close the subfilter stress and scalar flux terms. Solutions to the nonpremixed steady flamelet equations are precomputed using FlameMaster~\cite{pitsch1998flamemaster} and stored in the RFPV thermochemical database. The gas-phase kinetic mechanism of Refs.~\cite{blanquart2009chemical, narayanaswamy2010consistent} is considered, consisting of 1804 reactions and 158 species, which includes PAH species up to four aromatic rings. The soot model considers 12 soot sections so 36 total soot scalars (in addition to the square of the total number density to obtain the subfilter intermittency), and the section size distribution considers a spacing factor (ratio between two consecutive section volume sizes) of 7, to accurately track the soot size distribution and focus on a reasonable soot size region, based on Ref.~\cite{colman2023BMMSM}.

\section{Results\label{sec:results}} \addvspace{10pt}

%%% ATTENTION
\begin{figure*}[h!]
\centering
\vspace{-0.4 in}
\includegraphics[trim={0 4mm 0 0},clip,width=133mm]{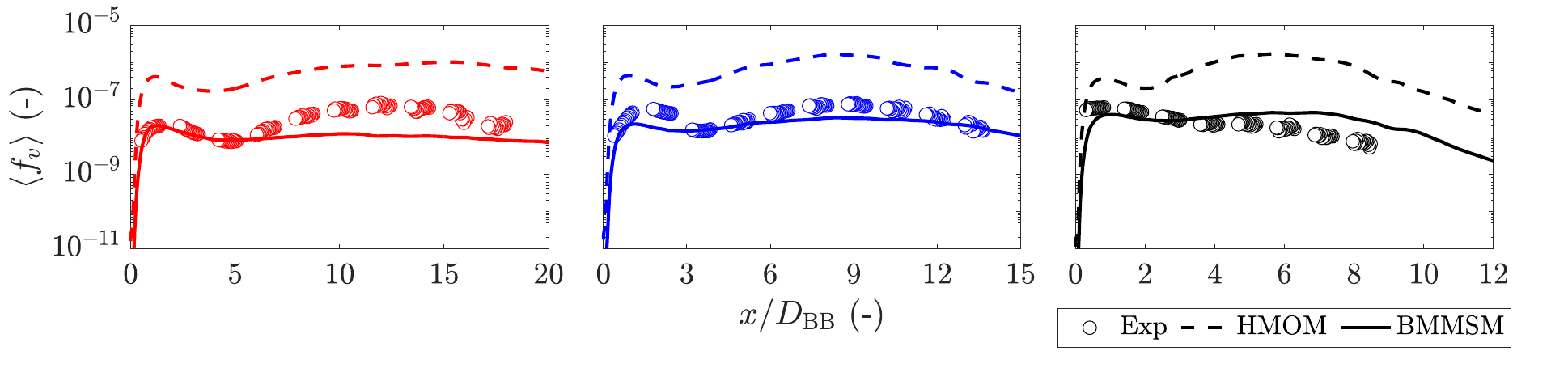}\vspace{5 pt}
\caption{\footnotesize Centerline profile of the mean soot volume fraction in the flame series: ENB-1 (left, red), ENB-2 (middle, blue), and ENB-3 (right, black). Symbols: experimental measurements. Lines: LES results using HMOM (dashed) and BMMSM (solid).}
\label{fig:fv_CL_3_HB}
\end{figure*}

%OLD CAPTION Centerline profile of the mean soot volume fraction in the flame series: ENB-1 (left, red), ENB-2 (middle, blue), and ENB-3 (right, black). Experimental measurements are indicated with symbols and LES results using HMOM (dashed) and BMMSM (solid) with lines.
%%%

In this section, results are presented in three sections. First, computational results using BMMSM are validated against experimental measurements of soot volume fraction \cite{rowhani2021soot}, which are also compared against computational results using HMOM \cite{colman2023largeBBF}. Second, the soot size distribution results using BMMSM are presented, highlighting the influence of the bluff body diameter. Third, an analysis of fields of source terms is performed to understand the impact of soot statistical model on soot evolution in the different regions of the bluff body flame.

\subsection{Soot volume fraction\label{subsec:soot_volume_fraction}} \addvspace{10pt}

Mean soot volume fraction profiles along the centerline are plotted in Fig.~\ref{fig:fv_CL_3_HB}. Computational results using BMMSM and HMOM are compared with experimental measurements. The three regions of the flame are well distinguished. First, the recirculation zone (0$\lesssim x/\DBB\lesssim1.6$) \cite{rowhani2021soot} allows for long residence times, so a first peak of soot volume fraction appears, both of them increasing with the bluff body diameter. Second, the neck region (1.6$\lesssim x/\DBB\lesssim2.2$) \cite{rowhani2021soot}, which is characterized by a highly strained flow and only soot escaping from the recirculation zone and passively transported downstream \cite{colman2023largeBBF}, exhibits a decrease in the soot volume fraction. Finally, the jet-like region ($x/\DBB\gtrsim2.2$) \cite{rowhani2021soot}, in which soot grows, exhibits a second peak of soot volume fraction that decreases as the bluff body diameter increases. Both BMMSM and HMOM are capable of reproducing the qualitative soot volume fraction trends as in the experiments. However, HMOM overpredicts the experimental measurements by about one order of magnitude in the first two flames and up to two orders of magnitude in ENB-3 ($x/\DBB\approx6$). On the other hand, BMMSM introduces a remarkable improvement in soot volume fraction predictions in the entire flame across the series. Some discrepancies do remain with BMMSM, mostly in the downstream jet-like region with an underprediction by a factor of up to six in ENB-1 and an overprediction by a factor of up to four in ENB-3.

\begin{figure*}[t!]
\centering
\vspace{-0.4 in}
\includegraphics[width=133mm]{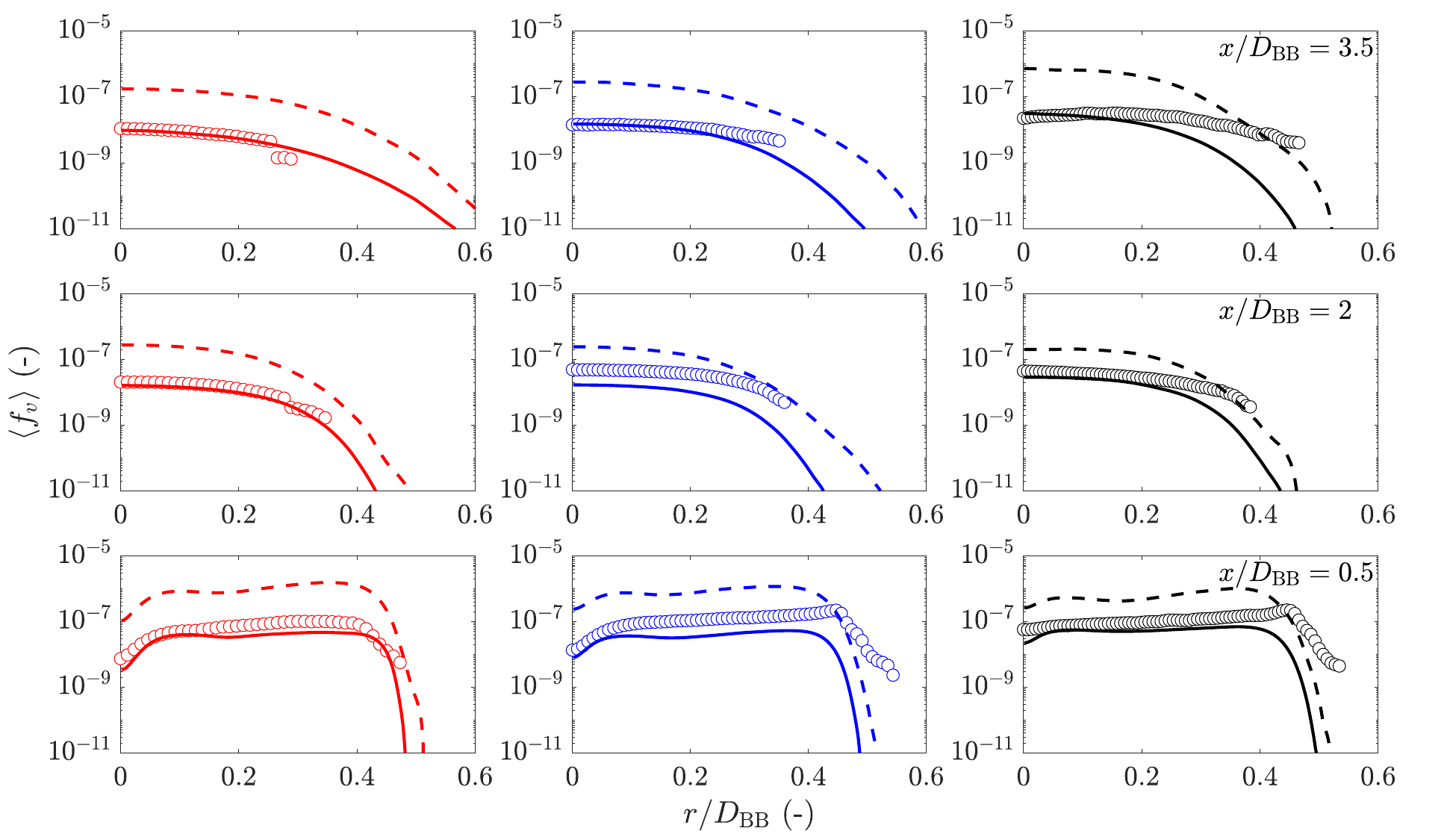}\vspace{5 pt}
\caption{\footnotesize Radial profiles of the mean soot volume fraction in the flame series at three streamwise locations: $x/\DBB=3.5$ (top), $x/\DBB=2$ (middle), and $x/\DBB=0.5$ (bottom). Lines styles are the same as those from Fig.~\ref{fig:fv_CL_3_HB}.}
\label{fig:fv_radial_3_HB}
\end{figure*}
%%%

Radial profiles of mean soot volume fraction profiles are plotted in Fig.~\ref{fig:fv_radial_3_HB} for the three flames at three different axial locations: in the recirculation zone at $x/\DBB=0.5$, in the neck zone at $x/\DBB=2$, and in the jet-like region at $x/\DBB=3.5$.

In the recirculation zone, relative low soot is observed at the inner shear layer between the jet and the coflow, increases outwardly and peaks at $r/\DBB\approx0.5$, and finally decreases at the outer shear layer between the recirculation zone and the coflow. BMMSM predicts well the soot volume fraction up to $r/\DBB\approx0.4$ in all three flames compared to experiments. Soot volume fraction at inner shear layer increases with $\DBB$, and BMMSM is able to capture this trend. The uniform radial profile of soot volume fraction within the recirculation zone is improved using BMMSM compared to HMOM. Compared to the experiments in ENB-2, an underprediction in the order of factor 2 is obtained using BMMSM versus an overprediction of factor 8 using HMOM. The soot volume fraction decrease is too sharp at the outer shear layer using both BMMSM and HMOM, which is attributed as in previous works~\cite{mueller2013experimental,colman2023largeBBF} to using a bulk velocity profile for the coflow.

In the neck region, BMMSM results are in good agreement with experiments. In ENB-1 soot is accurately predicts the soot volume fraction. In ENB-2, soot volume fraction is underpredicted using BMMSM with a difference of factor 3 close to the centerline compared to experiments. In ENB-3 the soot volume fraction is again well predicted using BMMSM compared to experiments. Overall, the error observed in the outer shear layer from the recirculation zone, with large bluff body diameters, propagates downstream to the neck region, which explains the discrepancies of one order of magnitude in the outermost part of the soot volume fraction profile compared to experiments. Compared to the recirculation zone, a marked reduction of soot is observed, due to the high strain conditions distinctive of this region.

In the jet-like region, BMMSM performs in the same fashion as in the previous regions compared to the experiments, with very good agreement for ENB-1 and ENB-2. Computational results using BMMSM and HMOM reproduce same profile trends, with soot decreasing in the radial direction outwardly. A significant decrease of soot is obtained in the outer part of the flame as $\DBB$ increases. Additionally, soot increases as $\DBB$ increases, which is well captured by the computational models. Compared to the neck region, soot starts to grow after leaving the high-strain neck region, as expected, and the mechanisms that leads to these effects will be explained later.

Profiles of rms soot volume fraction fluctuations show similar behavior and are not shown for the sake of brevity. In general, the soot volume fraction structures are well reproduced by the computational models, with the same trends as presented before, with HMOM an order of magnitude above the experiments while BMMSM widely improves the predictions.

% \subsection{\hl{Soot intermittency}\label{subsec:soot_rint}} \addvspace{10pt}
% \hl{If space!}

\subsection{Soot size distribution\label{subsec:soot_size_distribution}} \addvspace{10pt}

Computational results for the particle size distribution function (PSDF) using BMMSM are analyzed in this section. The PSDF is defined as in previous work \cite{yang2019multi,colman2023BMMSM}.
% The PSDF is defined as in previous work \cite{colman2023BMMSM}:
% \begin{equation}
%     \hat{n}(d_{p})=\left.\frac{1}{N}\frac{{\rm d} N(d_{p})}{{\rm d} \log d_{p}}\right\rvert_{d_{p}},
%     \label{eq:n_hat_def}
% \end{equation}
% \noindent
% where $d_{p}$ is the particle diameter, $N(d_{p})$ is the reconstructed soot particle cumulative distribution function (CDF), and $N=N(+\infty)$ is the total particle number density.
No experimental measurements of the PSDF exist for this configuration, so the PSDF ($\hat{n}(d_{p})$) is computed for particle diameters $d_{p}>2\rm\:nm$ to mimic instrument limitations as in previous work~\cite{colman2023BMMSM}.

%%%
\begin{figure*}[t!]
\centering
\vspace{-0.4 in}
\includegraphics[width=130mm]{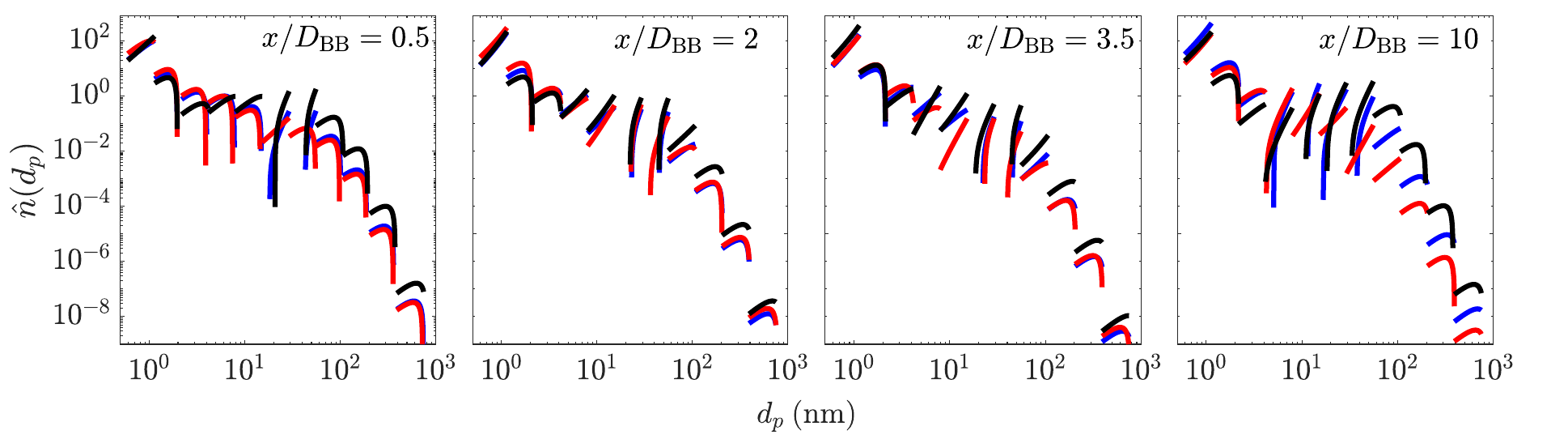}\vspace{2 pt}
\caption{\footnotesize PSDF at several locations along the centerline in the flame series at four different locations. Lines styles are the same as those from Fig.~\ref{fig:fv_CL_3_HB}.}
\label{fig:PSD_CL_3}
\end{figure*}
%%%

%%%
\begin{figure*}[t!]
\centering
% \vspace{-0.4 in}
\includegraphics[width=130mm]{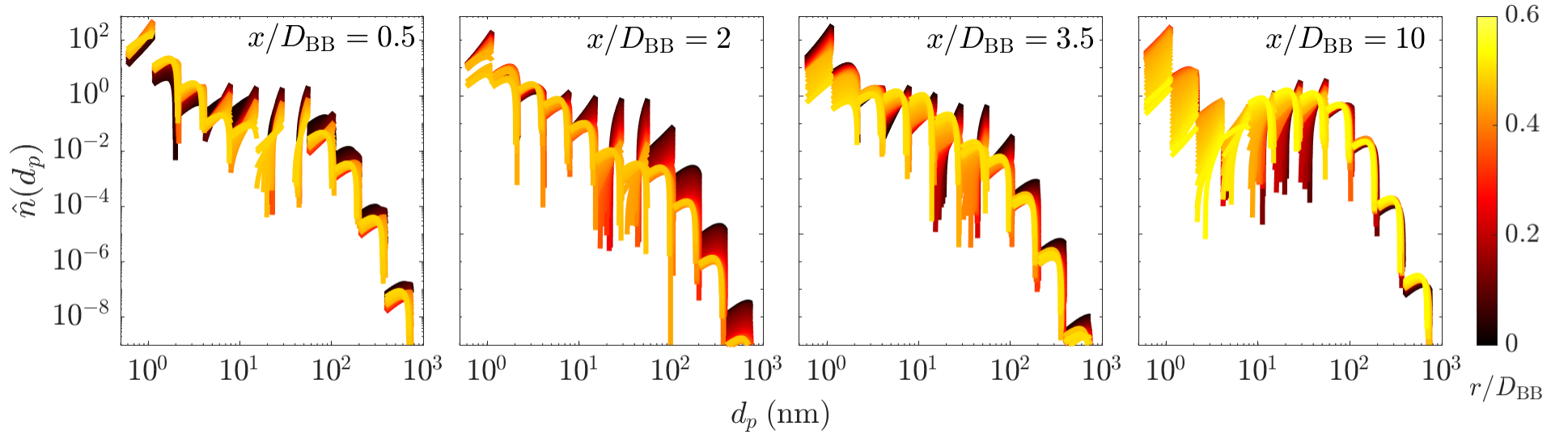}\vspace{2 pt}
%\vspace{10 pt}
\caption{\footnotesize PSDF evolution in the radial direction in ENB-3 at four different locations. The colorbar indicates the radial distance with respect to the centerline.}
\label{fig:PSD_rad_ENB3}
\end{figure*}
%%%

The PSDF as a function of the particle diameter is plotted in Fig.~\ref{fig:PSD_CL_3} for the three flames at four different locations along the centerline: in the recirculation zone at $x/\DBB=0.5$, in the neck region at $x/\DBB=2$, and in the jet-like region at $x/\DBB=3.5$ and $x/\DBB=10$. The PSDF exhibits a bimodal shape as the bluff body diameter increases, in the recirculation zone and in the most downstream part of the jet-like region ($x/\DBB=10$), due to long residence times, which increase with bluff body diameter. Sightly larger particles at the recirculation zone ($x/\DBB=0.5$) are observed in ENB-3, with a distinct second peak at $d_{p}\approx100\rm\:nm$. These large particles disappear in the neck region ($x/\DBB=2$) and the size distributions in the neck and near jet-like region ($x/\DBB=3.5$) are quite similar across the flame series. In the most downstream location, soot size increases, revealing a more pronounced second peak as the $\DBB$ increases. Overall, medium-sized particles ($20\rm\:nm$ to $100\rm\:nm$) show an almost uniform PSDF that decreases in magnitude in the first three regions, due to the highly-strained environment in the neck region. This is different to what is observed in a turbulent jet flame \cite{colman2023BMMSM}, where the upstream size distribution is strictly unimodal. Indeed, since the soot particles grow in the recirculation zone, soot that escapes to the neck region will be of a larger size compared to particles evolving along a jet flame with relatively shorter residence times.

The evolution of the size distribution in the radial direction is plotted in Fig.~\ref{fig:PSD_rad_ENB3} for the same locations as before. The colorbar indicates the normalized radial location. Only ENB-3 results are shown. In the recirculation zone, the size distribution show an almost constant PSDF for small and larger particles, whereas a decreasing PSDF of medium-sized particles, radially outward up to $r/\DBB\approx0.5$. In the neck zone, the PSDF decreases at medium to large size particles radially outward up to $r/\DBB\approx0.4$. The neck results also indicate that larger particles are preferentially located closer to the centerline. In the early jet-like region, the PSDF decreases radially outward up to $r/\DBB\approx0.3$ and then increases as the number density decreases up to $r/\DBB\approx0.45$. In the most downstream part, the PSDF exhibits the second mode as the the radial location progresses outwardly, showing a trough that starts at $d_{p}\approx5\rm\:nm$ that shift to the left, which can be attributed to the oxidation, reaching the outermost location. Overall, the first three regions show similar PSDF as in advanced stages of a jet flame~\cite{colman2023BMMSM}, but in a more compact space as residence times are much longer due to the recirculation.

\subsection{Source terms\label{subsec:soot_size_distribution}} \addvspace{10pt}

\begin{figure*}[t!]
\centering
\vspace{-0.4 in}
\includegraphics[width=144mm]{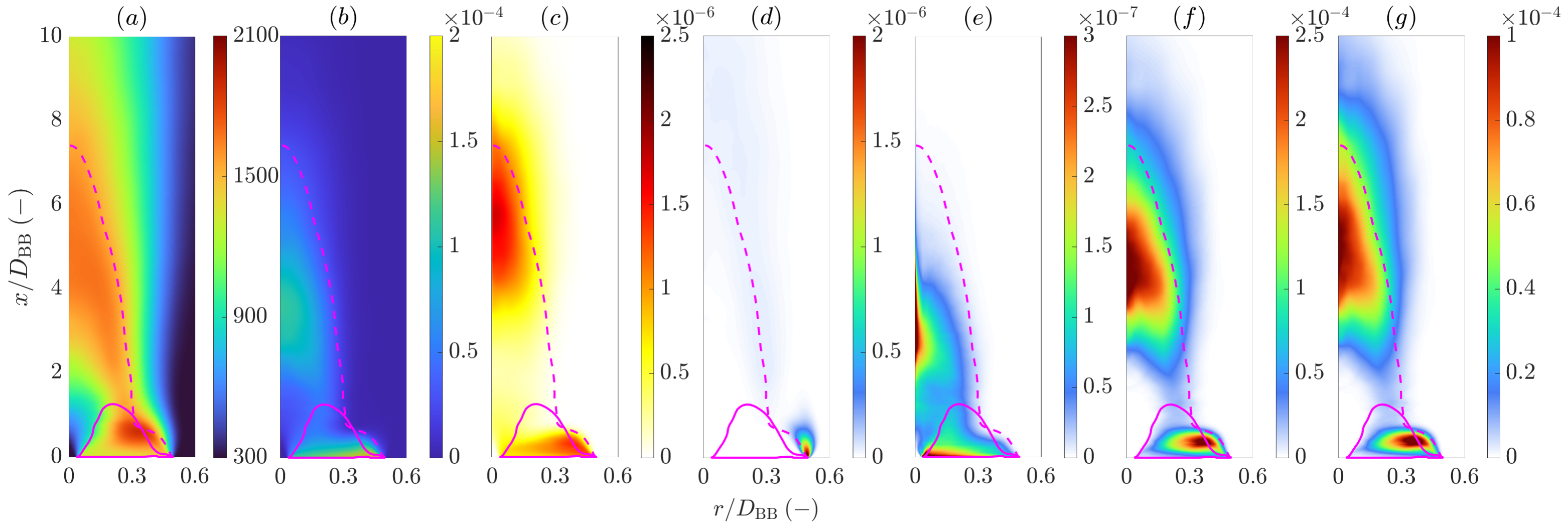}\vspace{5 pt}
\caption{\footnotesize Fields of mean soot-related quantities in ENB-3 using HMOM: $(a)$ temperature $T\rm\:[K]$; $(b)$ PAH mass fraction $Y_{\rm PAH}\rm\:[-]$; $(c)$ soot volume fraction $f_{v}\rm\:[-]$; and soot volume fraction source terms $df_{v}/dt\rm\:[s^{-1}]$ for $(d)$ nucleation, $(e)$ condensation, $(f)$ surface growth, and $(g)$ oxidation (magnitude). The solid line corresponds to the zero mean axial velocity iso-contour, and the dashed line corresponds to the mean stoichiometric mixture fraction iso-countour.}
\label{fig:STfv_ENB3_HMOM}
\end{figure*}

\begin{figure*}[t!]
\centering
% \vspace{-0.4 in}
\includegraphics[width=144mm]{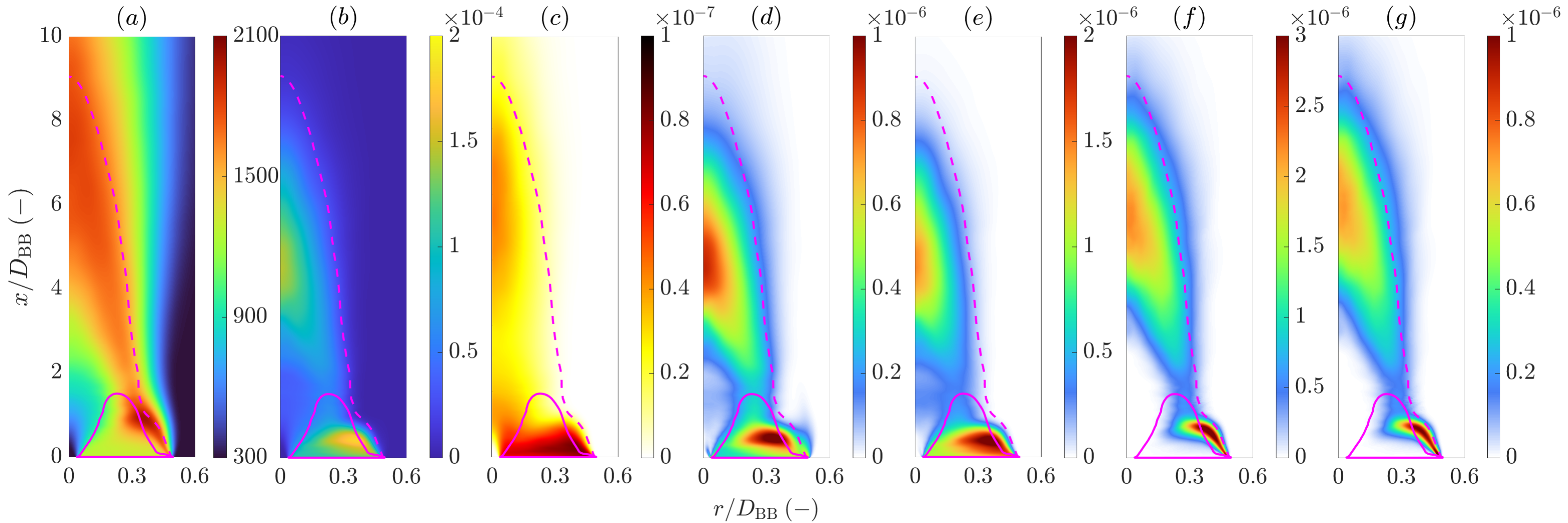}\vspace{5 pt}
\caption{\footnotesize Fields of mean soot-related quantities in ENB-3 using BMMSM as shown in Fig.~\ref{fig:STfv_ENB3_HMOM}.}
\label{fig:STfv_ENB3_BMMSM}
\end{figure*}

An extensive analysis of the soot source terms is carried out to reveal the underlying mechanisms of the evolution of the soot size distribution in the bluff body flame and how the choice of the soot statistical mechanism has an influence on global soot results. As in the previous section, only case ENB-3 is shown, and similar conclusions are obtained by analyzing the other flames. Fields of mean soot-related quantities, such as temperature, PAH mass fraction, and soot volume fraction $f_{v}$ and its source terms $df_{v}/dt$ are computed for both BMMSM and HMOM \cite{colman2023BMMSM}, which are plotted in Figs.~\ref{fig:STfv_ENB3_HMOM} and \ref{fig:STfv_ENB3_BMMSM}, respectively. The plots also include: a continuous line indicating an approximate of the mean recirculation zone, determined as the iso-contour of zero mean axial velocity, and a dashed line representing the mean flame front, obtained by computing the iso-contour of mean stoichiometric mixture fraction. These lines imply that both the length flame and the recirculation zone height are longer using BMMSM compared to HMOM. Compared to experimental measurements \cite{rowhani2022effects}, the recirculation zone height using BMMSM is about 17\% shorter whereas HMOM doubles this difference.

%%%
%%%
%%%
%%%

The mean temperature field is shown in column $(a)$. The temperature using HMOM is significantly lower than that using BMMSM, since the soot volume fraction, shown in column $(c)$, in HMOM is much higher than that of BMMSM, resulting in high radiation losses. The low temperature impacts the production of PAH, so low PAH mass fraction, shown in column $(b)$, is observed, especially in the recirculation zone and the jet-like region.

This fact reveals a remarkably outcome: the affected PAH formation using HMOM severely penalizes the nucleation rates, shown in column $(d)$, in the entire flame and is limited only in the outer shear layer. On the other hand, with BMMSM, nucleation occurs throughout the recirculation zone and even in the jet-like region. The condensation rate, shown in column $(e)$, is also affected for the same reason, since both are function of the square of the PAH mass fraction. This means that the soot escaping from the recirculation zone is only passively transported in the most highly-strained part of the neck zone. Soot nucleates and condenses in both the outer part of the neck region and in the jet-like region, where the temperature is sufficiently high. With HMOM, no new soot forms downstream of the recirculation zone. Finally, since soot surface growth rates in column $(f)$ and oxidation rates in column $(g)$ scale with the local amount of soot, HMOM results are about two orders of magnitude higher than those of BMMSM. In BMMSM, fields of both surface growth and oxidation rate are more confined to fuel-rich regions compared to HMOM as also observed in a turbulent nonpremixed jet flame \cite{colman2023BMMSM}.
 
These mechanisms also explain how the size distribution evolves (Fig.~\ref{fig:PSD_rad_ENB3}).  In the recirculation zone, medium sized particles are mostly obtained, as these coagulate and oxidize, and more small particles survive, reaching the outer shear layer as they nucleate and oxidize (with similar rates). In the neck region, the PSDF decreases radially outward for the small particles because of nucleation and oxidation and for the larger particles because of coagulation and oxidation. In the jet-like region, as the nucleated and escaping particles move downstream of the flame or radially outward, they coagulate and oxidize, resulting in bimodal distributions for longer residence times.

%This elucidates the importance on the choice of the soot statistical model, highlighting its role to reproduce soot volume fraction and size distribution, as well as to reveal fundamental mechanisms in turbulent sooting flames.

Interestingly, even the global soot quantities are highly sensitive to consideration of the size distribution in these bluff body flames. As discussed in previous work with BMMSM in a turbulent nonpremixed flames, HMOM and BMMSM have slightly different surface growth and oxidation distributions since the latter has a richer description of the range of particle sizes. In the jet flame, the influence of these differences are modest, but the bluff body flames are much more sensitive to these smaller differences. This reveals the importance of the soot statistical model and the need to have computationally efficient statistical models for the size distribution that are not much more expensive than methods of moments \cite{colman2023BMMSM} even for predicting global soot quantities and fundamental soot dynamics in turbulent soot flames.

\section{Conclusions\label{sec:conclusions}} \addvspace{10pt}

% The Declaration of competing interest, Acknowledgements, Supplementary material (if included), and References section headings are not numbered. The font and spacing are the same as those for regular section headings.

A turbulent nonpremixed bluff body flame series was studied using LES and the new BMMSM model for soot to investigate the evolution of soot size distribution. Computational results using BMMSM were first validated against experimental measurements and compared to computational results from previous work using HMOM. Significant improvement was observed in the prediction of soot volume fraction compared to this previous model, and very good agreement with the experimental measurements was achieved. Computational results using BMMSM reproduced favorably the soot trends and quantities across the flame series. The differences between BMMSM and HMOM results confirm the observations of previous work in a nonpremixed jet flame \cite{colman2023BMMSM}. However, these bluff body flames are much more sensitive to differences in the statistical models than the nonpremixed jet flame.

%Small discrepancies were noticed using both computational models in the outer shear layer, compared to the experimental results, which propagates to downstream the fuel-jet flame flame and are attributed to the coflow inlet velocity boundary conditions, based on past work. Further investigation is required to answer this question and are left for future work.

The evolution of the soot size distribution was investigated computationally in the three flames. The PSDF in the different flame regions is less sensitive to the bluff body diameter compared to the soot volume fraction. The flame with the largest diameter exhibited a slightly more pronounced bimodality in both the recirculation zone and the jet-like region. The presence of a large number of medium-sized particles indicates that the residence time has significant influence in the PSDF, featuring a more uniform PSDF in this size interval. This is different from previous work in turbulent nonpremixed jet flames \cite{colman2023BMMSM}, where the PSDF is strictly unimodal in upstream regions of the flame. The bluff body flame presented rather similar characteristics to the most downstream part of the jet flame, which includes bimodal distributions.

The mechanism of the evolution of soot size distribution was analyzed. Fields of soot related quantities revealed different soot behavior across the flame using BMMSM and compared to HMOM. The overprediction of soot using HMOM inhibits soot formation and condensation in the recirculation zone and jet-like regions. Conversely, BMMSM not only shows that soot forms and grows via condensation in those regions but also that these quantities are significant in the entire flame compared to surface growth and oxidation. Additionally, BMMSM shows that the soot escaping from the recirculation zone to the neck region is transported passively only in the inner part of highly-strained region. These mechanism also supports the evolution of the soot size distribution in the entire flame. These conclusions strengthen the concept of accurate soot size distribution models to both reproduce global soot quantities accurately and reveal fundamental mechanisms in turbulent sooting flames.

% \section{Nomenclature and appendix\label{sec:nomapp}} \addvspace{10pt}

% A nomenclature section and appendix are rarely included in {\em Proceedings} papers, because of length limitations. In the event that a nomenclature section is included, it should appear before the first numbered section heading under the unnumbered section heading “Nomenclature.” 
 
% If an appendix in included, it should appear immediately before the references under the unnumbered section heading ``Appendix.''

% The formatting for these section headings should be the same as that used for the Declaration of competing interest, Acknowledgements, Supplementary material, and References.

\acknowledgement{Declaration of competing interest} \addvspace{10pt}

% This is a required section, in which authors are to disclose any competing interests relating to the work reported in the paper. See \cite{elsevier_ci} for Elsevier's policy. Authors will be required to complete a form related to competing interests at the time of manuscript submission. An appropriate statement in the case of no competing interests to report is given in the following paragraph.

The authors declare that they have no known competing financial interests or personal relationships that could have appeared to influence the work reported in this paper.

% Use the acknowledgement environment defined in the template for this section, not \verb+\section*+. 

\acknowledgement{Acknowledgments} \addvspace{10pt}

The authors gratefully acknowledge funding from the National Science Foundation, Award CBET-2028318. The simulations presented in this article were performed on computational resources supported by the Princeton Institute for Computational Science and Engineering (PICSciE) and the Princeton University Office of Information Technology’s Research Computing department.

%Use the acknowledgement environment defined in the template for this section, not \verb+\section*+. 

% \acknowledgement{Supplementary material} \addvspace{10pt}

%If supplementary material is submitted along with the manuscript, that should be noted here. In the event that the manuscript is accepted for publication in the {\em Proceedings}, a DOI link to the online supplementary material will be included in the published paper.

% Use the acknowledgement environment defined in the template for this section, not \verb+\section*+. 

% -------------------------------------------------------------------- %
% -------------------------------------------------------------------- %
% -------------------------------------------------------------------- %

 \footnotesize
 \baselineskip 9pt

% -------------------------------------------------------------------- %
% -------------------------------------------------------------------- %
% -------------------------------------------------------------------- %

\bibliographystyle{pci}
\bibliography{BMMSM_BBF_paper}

% -------------------------------------------------------------------- %
% -------------------------------------------------------------------- %
% -------------------------------------------------------------------- %

\newpage

\small
\baselineskip 10pt

% -------------------------------------------------------------------- %
% -------------------------------------------------------------------- %
% -------------------------------------------------------------------- %

% -------------------------------------------------------------------- %
% -------------------------------------------------------------------- %
% -------------------------------------------------------------------- %

\end{document}